\documentclass[11pt]{article}
\usepackage[left=3cm, right=3cm, top=3cm, bottom=3cm]{geometry}
\usepackage{amsmath, amsthm, amssymb}
\usepackage[osf]{CormorantGaramond}
\usepackage{graphicx}
\usepackage{todonotes}
\usepackage{enumitem}

\newcommand{\ie}{\textit{i.e.,\ }}
\newcommand{\eg}{\textit{e.g.,\ }}

\renewcommand{\S}{\ensuremath{\mathbb{S}}}
\newcommand{\midi}{\textsc{Midi}}

\newcommand{\dancing}{Dancing\,\midi}
\newcommand{\fl}{\ensuremath{\flat}}

\DeclareMathOperator{\splitseq}{\textsc{\small split}}
\DeclareMathOperator{\concatseq}{\textsc{\small concat}}
\newcommand{\conc}{$\concatseq$}
\newcommand{\spl}{$\splitseq$}

\usepackage{enumitem}

\usepackage{algorithm}
\usepackage[noend]{algpseudocode}

\theoremstyle{definition}

\title{Smart Edition of \midi\ Files}
\author{Pierre Roy  and Fran\c{c}ois Pachet \\ Spotify}
\date{}

\begin{document}

\maketitle

\begin{abstract} We address the issue of editing musical performance data, in particular \midi\ files representing human musical performances.
Editing such sequences raises specific issues due to the ambiguous nature of musical objects.
The first source of ambiguity is that musicians naturally produce many deviations from the metrical frame. These deviations may be intentional or subconscious, but they play an important role in conveying the groove or feeling of a performance.
Relations between musical elements are also usually implicit, creating even more ambiguity. A note is in relation with the surrounding notes in many possible ways: it can be part of a melodic pattern, it can also play a harmonic role with the simultaneous notes, or be a pedal-tone.
All these aspects play an essential role that should be preserved, as much as possible, when editing musical sequences.

In this paper, we contribute specifically to the problem of editing non-quantized, metrical musical sequences represented as \midi\ files. We first list of number of problems caused by the use of naive edition operations applied to performance data, using a motivating example.
We then introduce a model, called Dancing \midi, based on 1) two desirable, well-defined properties for edit operations and 2) two well-defined operations, \spl\ and \conc, with an implementation.
We show that our model formally satisfies the two properties, and that it prevents most of the problems that occur with naive edit operations on our motivating example, as well as on a real-world example using an automatic harmonizer.
\end{abstract}

\section{Introduction}\label{sec:intro}


The term music performance denotes all musical artefacts produced by one or more human musicians playing music, such as a pianist performing a score or accompanying a singer, a violin section playing an orchestration of a piece or a jazz musician improvizing a solo on a given lead sheet.
Music performance can be represented in various ways, depending on the context of use: printed notation, such as scores or lead sheets, audio signals, or performance acquisition data, such as piano-rolls or \midi\ files. Each of these representations captures partial information about the music that is useful in certain contexts, with its own limitations \cite{Dannenberg:1993:CMJ:Brief}.
Printed notation offers information about the musical meaning of a piece, with explicit note names and chord labels (in, \eg lead sheets), and precise metrical and structural information, but it tells little about the sound.
Audio recordings render timbre and expression accurately, but provide no information about the score.
Symbolic representations of musical performance, such as \midi, provide precise timings are are therefore well adapted to edit operations, either by humans or by software.

The need for editing musical performance data arises from two situations. First, musicians often need to edit performance data when producing a new piece of music. For instance, a jazz pianist may play an improvized version of a song, but this improvization should be edited to accommodate for \textit{a posteriori} changes in the structure of the song. The second need comes from the rise of AI-based automatic music generation tools. These tools usually work by analyzing existing human performance data to produce new ones (see, e.g. \cite{briotdeep} for a survey). Whatever the algorithm used for learning and generating music, these tools call for editing means that preserve as far as possible the expressiveness of original sources.
We address the issue of editing musical performance data represented as \midi\ files, while preserving as much as possible its semantics, in a sense defined below.

However, editing music performance data raises specific issues related to the ambiguous nature of musical objects. The first source of ambiguity is that musicians produce many temporal deviations from the metrical frame. These deviations may be intentional or subconscious, but they play an important part in conveying the groove or feeling of a performance. Relations between musical elements are also usually implicit, creating even more ambiguity. A note is in relation with the surrounding notes in many possible ways: it can be part of a melodic pattern, it can also play a harmonic role with the simultaneous notes, or be a pedal-tone. All these aspects, although not explicitly represented in a \midi\ file, play an essential role that should be preserved, as much as possible, when editing such musical sequences.

The \midi\ format is widespread in the instrument industry and \midi\ editors are commonplace, for instance in Digital Audio Workstations. Paradoxically, the problem of editing \midi\ with \emph{semantic-preserving} operations was not  addressed yet, to our knowledge.
Attempts to provide semantically-preserving edit operations have been made on the audio domain (e.g. \cite{whittaker2004semantic}) but these are not transferrable to music performance data, as we explain below.

In human-computer interaction, \emph{cut}, \emph{copy} and \emph{paste} \cite{Tesler:2012:PHM:2212877.2212896} are the Holy Trinity of data manipulation. These three commands proved to be so useful that they are now incorporated in virtually every software, such as word processing, programming environments, graphics creation, photography, audio signal, or movie editing tools.
Recently, they have been extended to run across devices, enabling moving text or media from, for instance, a smartphone to a computer. These operations are simple and have a clear, unambiguous semantics: \emph{cut}, for instance, consists in selecting some data, say a word in a text, removing it from the text, and saving it to a clipboard for later use.

Each type of data to be edited raises its own editing issues that led to the development of specific editing techniques. For instance, edits of audio signals usually require cross fades to prevent clicks. Similarly, in movie editing, fade-in and fade-out are used to prevent harsh transitions in the image flow. Edge detection algorithms were developed to simplify object selection in image editing.
The case of \midi\ data is no exception. Every note in a musical work is related to the preceding, succeeding, and simultaneous notes in the piece. Moreover, every note is related to the metrical structure of the music.
In Section~\ref{sec:motivating-example}, we list a number of issues occurring when applying na\"ive edition commands to a musical stream.

In this paper, we restrict ourself to a specific type of musical performance data: non-quantized, metrical music data, i.e. performances which are recorded with free expression but with a fixed, known tempo.
This includes most of MIDI files available on the web (for instance \cite{Lakh} ). This excludes MIDI files consisting of free improvization, or music performance with no fixed tempo. Note that these could also be included in the scope of our system, using automatic downbeat estimation methods, but we do not consider this case in this paper.

It is not possible, to our knowledge, to define a precise semantics to musical performance in general.
In this paper we contribute to the problem of editing non-quantized, metrical musical sequences represented as \midi\ files in the following way:
\begin{enumerate}
\item We list of number of problems caused by the use of naive edition operations applied to performance data, using a motivating example;
\item We then introduce a model, called \dancing, based on 1) two desirable, well-defined properties for edit operations and 2) two well-defined operations, \spl\ and \conc, with an implementation. These primitives can be used to create higher-level operations, such as \emph{cut}, \emph{copy}, or \emph{paste};
\item We show that our model formally satisfies the two properties;
\item We show additionally that our model does not create most of the problems that occur with naive edit operations on our motivating example, as well as on a real-world example using an automatic harmonizer.
\end{enumerate}

\section{Motivating Example}
\label{sec:motivating-example}

\begin{figure}[htp!]\centering
    \includegraphics[width=.6\textwidth]{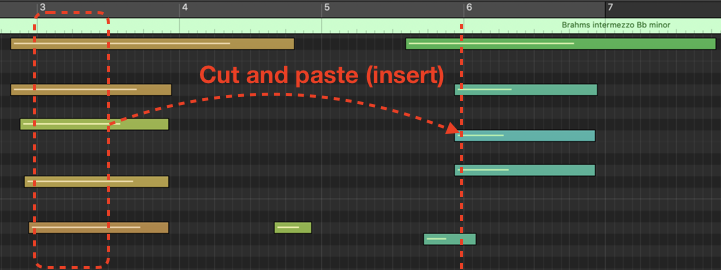}
    \caption{A piano roll with five measures extracted from a piece by Brahms. Colors indicate note velocities (blue is soft, green is medium, and brown is loud). A typical edit operation: the goal is to \emph{cut} the first two beats of Measure~3 and \emph{insert} them at the beginning of Measure~6.}
    \label{fig:motiv-ex-statement}
\end{figure}

\begin{figure}[htp!]\centering
    \includegraphics[width=.5\textwidth]{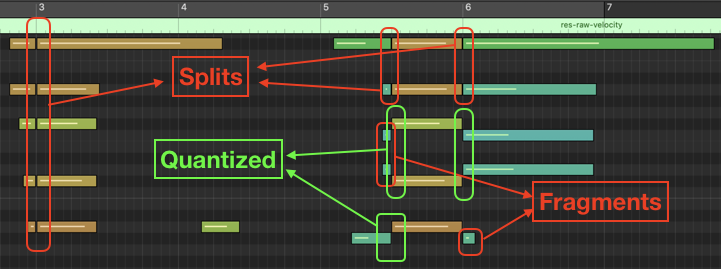}
    \caption{Raw editing the piano roll in Figure~\ref{fig:motiv-ex-statement} produces a poor musical result: long notes are split, residuals are created, some notes are quantized, and note velocities are inconsistent. This piano roll was obtained using Apple Logic Pro X \midi\ editor, using the ``split'' option, see Section~\ref{sec:state-of-the-art}.}
    \label{fig:motiv-ex-raw}
\end{figure}

Figure~\ref{fig:motiv-ex-statement} shows a piano roll representing five measures extracted from a \midi\ stream consisting of a performance capture of Johannes Brahms's Intermezzo in B\fl\ minor. Consider the problem of \emph{cutting} the first two beats of Measure~3 and \emph{inserting} these two beats at the beginning of Measure~6. Figure~\ref{fig:motiv-ex-raw} shows the piano roll produced when these operations are performed in a straightforward way, \ie when considering notes as mere time intervals. Notes that are played across the split temporal positions are segmented, leading to several musical inconsistencies. First, long notes, such as the highest notes, are split into several contiguous short notes. This alters the listening experience, as several attacks are heard, instead of a single one. Additionaly, the note \emph{velocities} (a \midi\ equivalent of loudness) are possibly changing at each new attack, which is unmusical. Another issue is that splitting notes with no consideration of the musical context leads to creating excessively short note fragments, which we call \emph{residuals}, \eg at the bottom right in Figure~\ref{fig:motiv-ex-raw}. Residuals are disturbing, especially if their velocity is high, and are somehow analogous to clicks in audio signals. Finally, a side-effect of this approach is that some notes are quantized (last two beats of Measure~5). As a result, slight temporal deviations present in the original \midi\ stream are lost in the process. Such temporal deviations are important parts of the performance, as they convey the groove, or feeling of the piece, as interpreted by the musician.

Here is a list of musical issues occurring when raw editing a \midi\ stream:
\begin{enumerate}
    \item Creation of \emph{residuals}, \ie excessively brief notes;
    \item Splitting long notes, creating superfluous attacks;
    \item Creating surprising, inconsistent changes in note velocities;
    \item Losing small temporal deviations with respect to the metrical structure, leading to unnecessary, undesirable quantization.
\end{enumerate}

\begin{figure}[htp!]\centering
    \includegraphics[width=.6\textwidth]{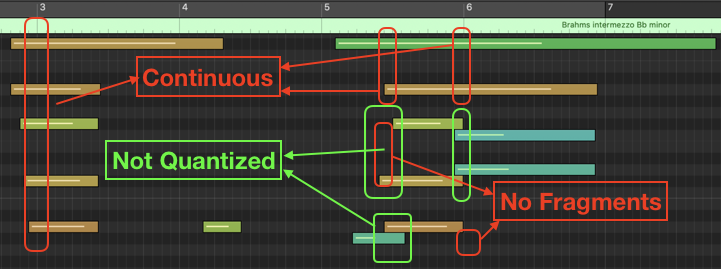}
    \caption{Solving the \midi\ edition problem stated in Figure~\ref{fig:motiv-ex-statement} using the model we present here. There are no short note residuals, long notes are held and there are no harsh changes in velocities. Small temporal deviations are preserved (no quantization). This is to be compared to Figure~\ref{fig:motiv-ex-raw}.}
    \label{fig:motiv-ex-dancing}
\end{figure}

Figure~\ref{fig:motiv-ex-dancing} shows another solution to the problem, obtained using the model presented in this article. Comparing Measure~5 in Figures~\ref{fig:motiv-ex-raw} and \ref{fig:motiv-ex-dancing} shows obvious differences in the two approaches: none of the issues produced with raw edits, shown in Figures~\ref{fig:motiv-ex-raw}, are present in the piano roll shown in Figures~\ref{fig:motiv-ex-dancing}.

\section{State of the Art}
\label{sec:state-of-the-art}

\begin{figure}[htp!]\centering
    \includegraphics[width=.6\textwidth]{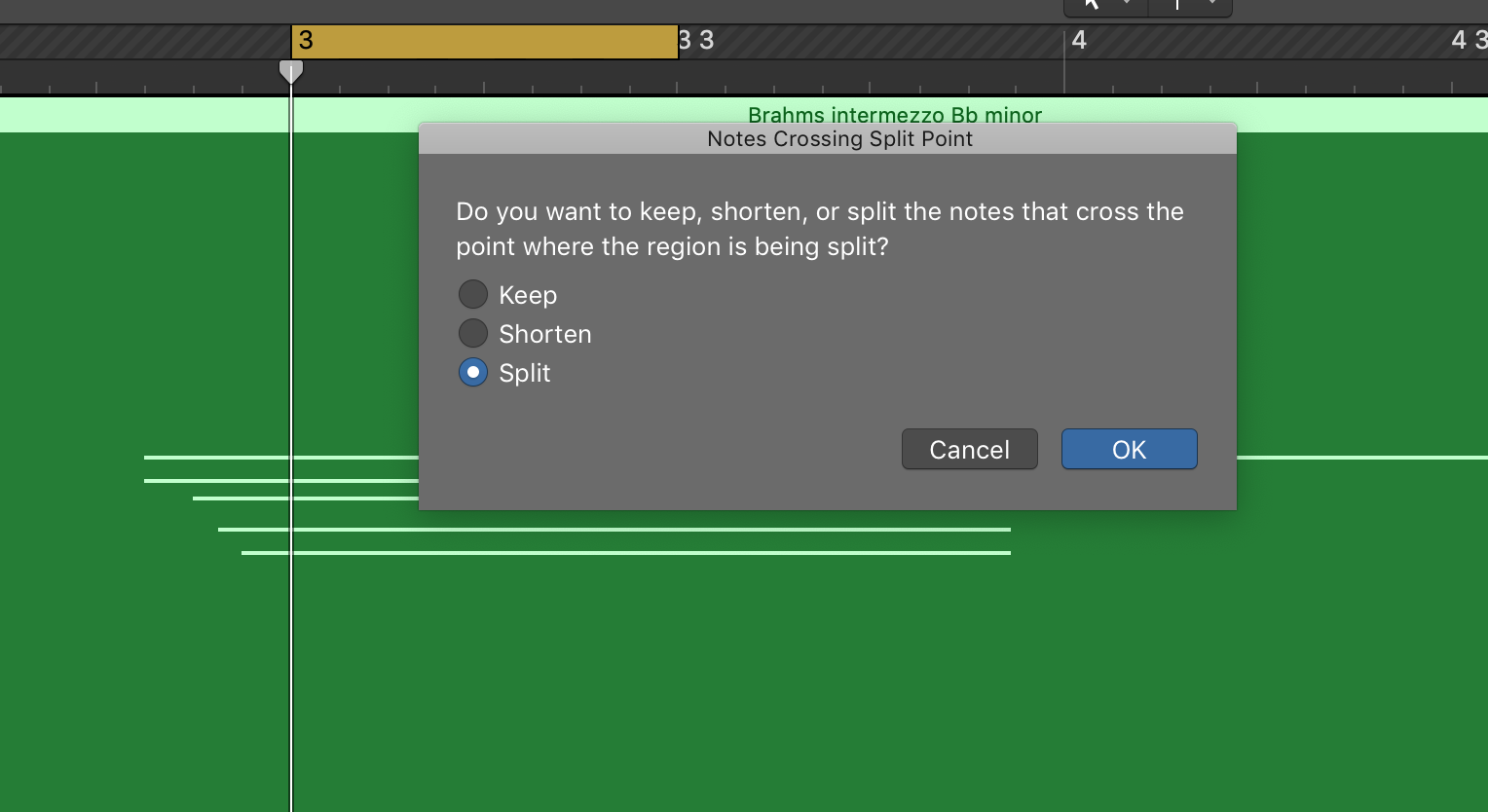}
    \caption{The menu that pops up when splitting a \midi\ stream with Apple Logic~Pro~X Digital Audio Workstation. Here, we split at both ends of a region covering the first two beats of Measure~3 (yellow zone). The system lets the user decide what they want explicitly regarding notes that cross the split region boundaries.}
    \label{fig:logic-split}
\end{figure}

LogicPro~X, the Digital Audio Workstation commercialized by Apple, features a full-featured \midi\ editor. As shown in Figure~\ref{fig:logic-split}, when editing a \midi\ stream, the decisions regarding notes that overlap with the selected regions are left to the user who has to decide whether to \emph{split}, \emph{shorten} (to fit the region boundaries) or \emph{keep} the notes. Figure~\ref{fig:motiv-ex-raw} shows the piano roll produced using the first option, split. In the latter case, keep, when pasting somewhere else, the notes will be put back with their original duration, even if it exceeds the region boundaries. This forces the user to decide explicitly what to do, and the decision applies to all notes, regardless of the musical context. Besides, this strategy leads to creating overlapping notes, which create ambiguous situations as the \midi\ format does not have a way to handle overlapping notes with the same pitch.

\begin{figure}[htp!]\centering
    \includegraphics[width=.6\textwidth]{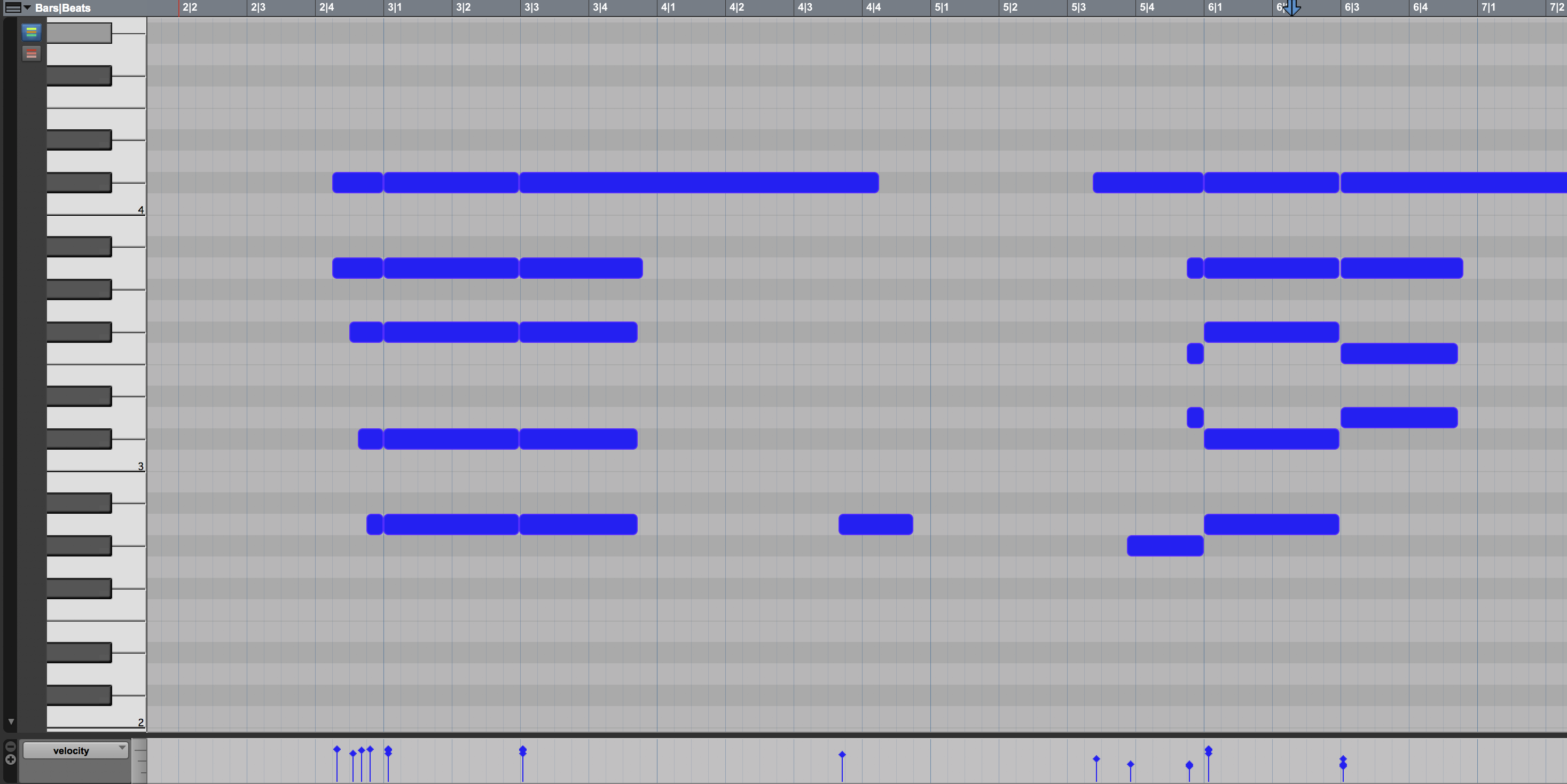}
    \caption{The piano roll produced when editing the motivating example using the piano roll module in Avid Pro Tools.}
    \label{fig:pro-tools-split}
\end{figure}

The piano roll panel in Avid Pro Tools, another major Digital Audio Workstation, offers less control on \midi\ edits than Logic Pro X. Figure~\ref{fig:pro-tools-split} shows the piano roll obtained with Pro Tools, when using the basic copy and paste functions on the motivating example. This piano roll is essentially the same as that in Figure~\ref{fig:motiv-ex-raw}, except note velocities are not displayed.

\section{The Model}\label{sec:model}

In this section, we present a model of temporal sequences that allows us to implement two primitives: \spl\ and \conc. The \spl\ primitive is used to break a \midi\ stream (or \midi\ file) at a specified temporal position, yielding two \midi\ streams: the first one contains the music played before the split position and the second one contains the music played after. The \conc\ operation takes two \midi\ streams as inputs and returns a single stream by appending the second stream to the first one. The model is called Dancing \midi\ since the underlying technique (see Section~\ref{sec:model-implementation}) bears some similarity with the idea of \emph{Dancing Links} that Donald Knuth developed in \cite{Knuth:2000:DancingLinks}

High-level edit operations, such as \emph{copy}, \emph{cut}, \emph{paste}, or \emph{insert}, may be performed by applying \spl\ and \conc\ to the right streams. For example, to \emph{cut} a \midi\ stream $S$ between temporal positions $t_1$ and $t_2$, we execute the sequence of primitive operations
\begin{enumerate}
    \item $(S', S_r) \gets \splitseq(S, t_2)$, \ie we split $S$ at $t_2$;
    \item $(S_l, S_m) \gets \splitseq(S', t_1)$, \ie we split $S'$ at $t_1$.
    \item Return $\concatseq(S_l, S_r)$;
\end{enumerate}
The stream $S_m$ is removed from  $S$ and it could be stored in a \midi\ clipboard for later use.
Similarly, to \emph{insert} a \midi\ stream $T$ in a stream $S$ at temporal position $t$, one can perform:
\begin{enumerate}
    \item $(S_1, S_2) \gets \splitseq(S, t)$, \ie split $S$ at $t$;
    \item Return $\concatseq\left(\concatseq\left(S_1, T\right), S_2\right)$.
\end{enumerate}

In our model, a \emph{continuous subdivision} of time coexists with a \emph{discrete, regular subdivision} of time, in, \eg beats or measures, which is equivalent to the metrical frame of a musical sequence. The constant distance between discrete temporal positions is denoted by $\delta$. Musical events, \eg notes, rests, chords, may occur at any continuous temporal position. On the contrary, \spl\ and \conc\ are applicable only on discrete temporal positions, \ie multiples of $\delta$. In short, musical events may be placed at any position, and take arbitrary durations, within a regularly decomposed time frame.

We made the practical choices to represent time by integers and to consider only sequences starting at time $0$ and whose duration is a multiple of the segmentation subdivision, which is denoted by $\delta$. These choices aim at simplifying the implementation and clarifying the presentation without limiting the generality of the model\footnote{In \midi\ files, time is represented by integer numbers, based on a predefined \emph{resolution}, typically 960 ticks per beats.}.

A \emph{(time) event} $e$ is defined by its start and end times, denoted by $e^-$ and $e^+$, two nonnegative integers, with $e^- \leq e^+$. We consider sequences of \emph{non-overlapping} time events. A sequence $S$ with a duration $d(S)$ is an ordered list of time events $E(S)=(e_1, \dots, e_n)$, such that:
    \begin{itemize}
        \item $d(S) \equiv 0 \pmod{\delta}$, \ie the duration of $S$ is a multiple of $\delta$,
        \item $e_n^+ \leq d(S)$, \ie all events are within the sequence, and
        \item $e_i^+ \leq e_{i+1}^-, \forall i = 1, \dots, n-1$, \ie there are no overlapping events.
    \end{itemize}
The set of all such sequences is denoted by $\S_\delta$.

The model handles sequences of non-overlapping time intervals (elements of $\S_\delta$), and therefore cannot directly deal with \midi\ streams, which contain overlapping intervals (\eg chords consist of three or more overlapping intervals). Therefore, we decompose a \midi\ stream into individual sequences of non-overlapping events, one for each unique pitch and unique \midi\ channel. We show, in Section~\ref{sec:model-on-piano-rolls}, how this approach applies to real \midi\ streams.

\subsection{Problem Statement}

We address the problem of implementing \spl\ and \conc\ in an \emph{efficient} and \emph{sensible} (from a musical viewpoint) way. The \spl\ primitive breaks a sequence at a specific temporal position and the \conc\ primitive returns a sequence formed by concatenating two sequences:
    \begin{align*}
      \splitseq\colon\S_\delta\times\{0,\delta,2\delta,\dots,d(S)\} &\to \S_\delta\times\S_\delta\\
                (S, t) &\mapsto (S_l, S_r),
    \end{align*}
where $t$ is the segmentation position, $S_l$ is the \emph{left} part of $S$, from $0$ to $t$, hence $d(S_l)=t$, and $S_r$ is the \emph{right} part of $S$, after position $t$, with $d(S_r)=d(S)-t$ and
    \begin{align*}
      \concatseq \colon \S_\delta\times\S_\delta &\to \S_\delta\\
                        (S_1, S_2) &\mapsto S = S_1 \oplus S_2.
    \end{align*}
where $S$ is a sequence of duration $d(S)=d(S_1)+d(S_2)$ constructed by appending $S_2$ to $S_1$.

The types above specify that \spl\ and \conc\ create sequences of $\S_\delta$, \ie, sequences with no overlapping events and with all events falling within the sequence's bounds.

We call \emph{residual} an event whose duration is shorter than a predefined threshold $\varepsilon$. We define two properties for \spl\ and \conc:
\begin{enumerate}[label=(\textbf{P\arabic*})]
    \item \label{prop:reversible} \spl\ and \conc\ are the \emph{inverse} of one another, i.e.,
    \begin{enumerate}[label=\alph*.]
        \item $\forall S\in\S_\delta, \forall t\in\{0,\delta, 2\delta, \dots,d(S)\}, \concatseq(\splitseq(S, t)) = S$\label{prop:reversible:a}
        \item $\forall S, T\in\S_\delta, \splitseq(\concatseq(S, T), d(S)) = (S, T)$\label{prop:reversible:b}
    \end{enumerate}
    \item \label{prop:residuals} \spl\ and \conc\ never \emph{create} residuals, i.e.,
    \begin{enumerate}[label=\alph*.]
        \item let $S\in\S_\delta, \forall t\in\{0,\delta, 2\delta, \dots,d(S)\}$ and $(S_1, S_2) = \splitseq(S, t)$, then $\forall e \in S_1 \cup S_2$,\\ $d(e) < \varepsilon \Rightarrow e \in S$,\label{prop:residuals:a}
        \item $\forall S, T\in\S_\delta, \forall e \in S \oplus T, d(e) < \varepsilon \Rightarrow e \in S \cup T$.\footnote{This is a simplification, as we explain in Section~\ref{sec:concat}}\label{prop:residuals:b}
    \end{enumerate}
\end{enumerate}

Property~\ref{prop:reversible} states that splitting a sequence and merging back the resulting sequences produces the original sequence and, conversely, concatenating two sequences and splitting them again at the same position, returns the two original sequences. This is to ensure that no information is lost upon splitting and concatenating sequences. Additionally, as we show in Section~\ref{sec:model-discussion}, this property offers the benefits of a powerful, generalized undo mechanism.

Property~\ref{prop:residuals} ensures that no residual is created: the only residuals appearing in a sequence obtained using \spl\ or \conc\ were already in the original sequence(s). Note that the second part, \ie \ref{prop:residuals}(\ref{prop:residuals:b}), is a bit simplified, as we explain in Section~\ref{sec:concat}.

It is easy to design \spl\ and \conc\ primitives that satisfy either \ref{prop:reversible} \textbf{or} \ref{prop:residuals}. However, as we will illustrate now, it is difficult to enforce \ref{prop:reversible} \textbf{and} \ref{prop:residuals}. It is the combination of these two properties that ensures that no information is lost and that no residual is created.

\begin{figure}[htp!]\centering
    \includegraphics[width=.6\columnwidth]{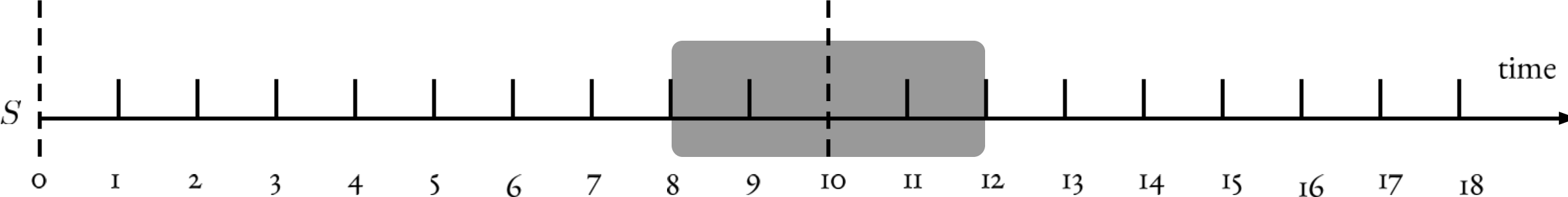}
    \caption{A simple sequence with a single event (gray box) and a regular subdivision of $10$ time units.}
    \label{fig:S}
\end{figure}

Figure~\ref{fig:S} shows a simple sequence $S$ of duration $d(S)=20$, with a regular temporal subdivision $\delta=10$ time units, and containing event $e$, with $e^-=8$ and $e^+=12$.
We consider the problem of splitting $S$ in its middle ($t=10$) and concatenating back the two extracted sequences.

\begin{figure}[htp!]\centering
    \includegraphics[width=.7\columnwidth]{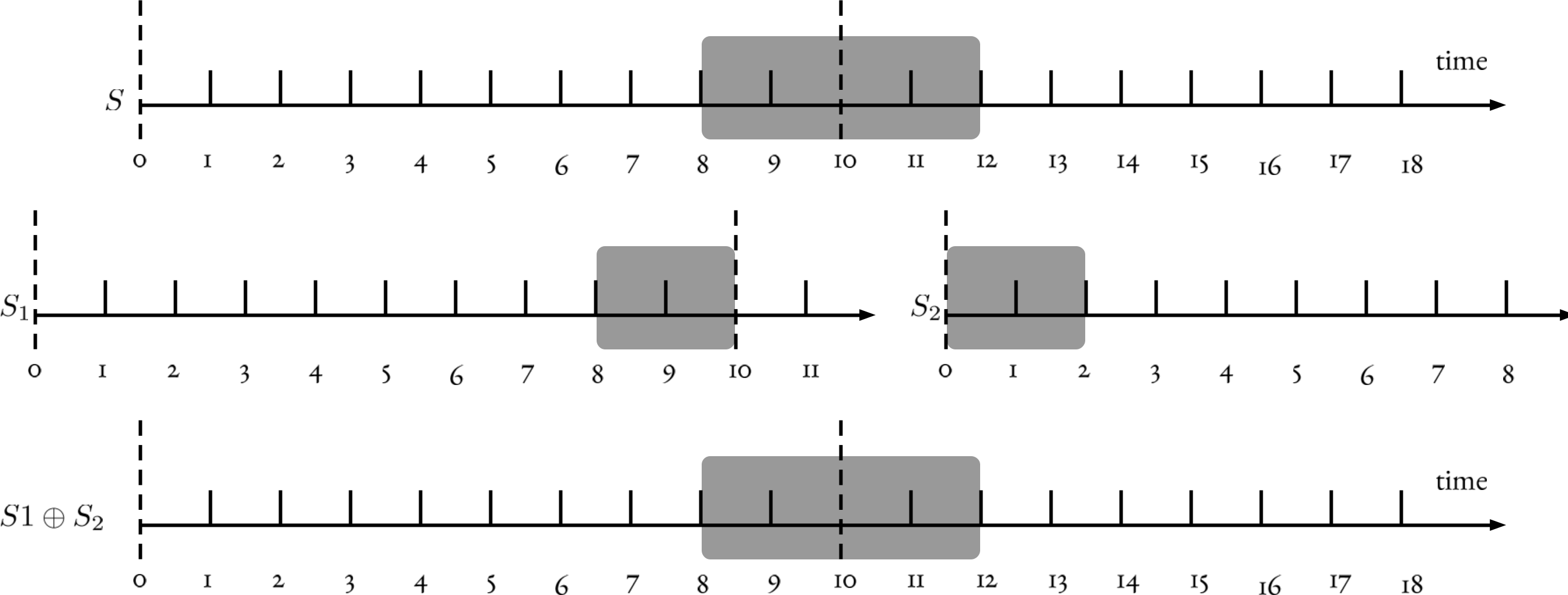}
    \caption{Straightforward approach: sequence $S$ (top) is split at $t=10$, yielding sequences $S_1$ and $S_2$ (middle), with two residuals (events of duration 2), then $S_1$ and $S_2$ are concatenated resulting in $S_1\oplus S_2$, which is identical to $S$, as expected.}
    \label{fig:split_S1_S2_S3_strict}
\end{figure}

Assuming the threshold for brief events is set to $\varepsilon=3$ time units, \spl\ and \conc\ should not create events shorter than $3$.
Figure~\ref{fig:split_S1_S2_S3_strict} shows the result of splitting $S$ into $S_1$ and $S_2$ and concatenating $S_1$ and $S_2$, performed in a straightforward way by raw event segmentation, \ie cut exactly at the specified position, and with no additional processing.
The concatenation $S_1 \oplus S_2$ is identical to the original sequence $S$, satisfying Property~\ref{prop:reversible}. However, this approach creates two events of duration $2$ in the split sequences $S_1$ and $S_2$, which violates the residual Property~\ref{prop:residuals} for $\varepsilon = 3$.

\begin{figure}[htp!]\centering
    \includegraphics[width=.7\columnwidth]{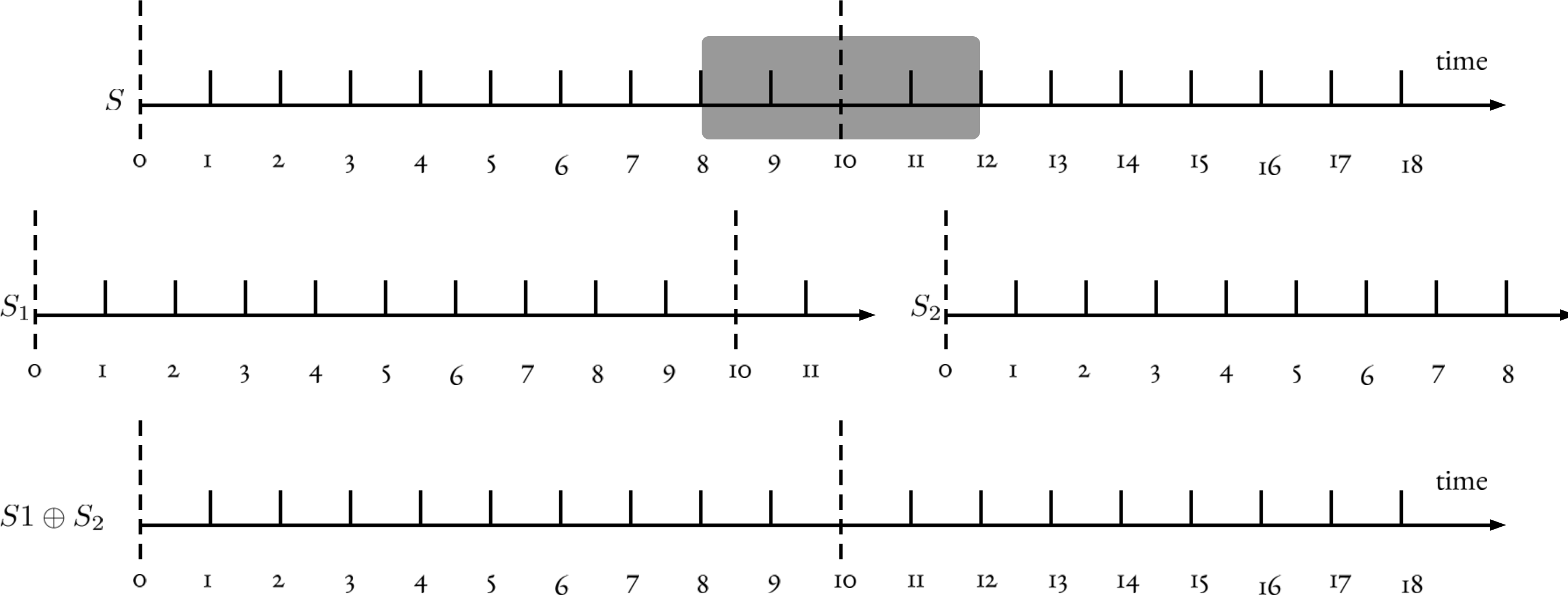}
    \caption{Applying the straightforward approach with a filter for residuals results in creating no residuals ($S_1$ and $S_2$ are empty, as expected), but violates the reversibility as $S_1 \oplus S_2 \neq S$.}
    \label{fig:split_S1_S2_S3_smart}
\end{figure}

On Figure~\ref{fig:split_S1_S2_S3_smart}, on the contrary, short fragments are omitted, by adding a simple filter, fulfilling Property~\ref{prop:residuals} as no residuals show up on split sequences $S_1$ and $S_2$. However, $S_1 \oplus S_2$ is empty, which violates the reversibility Property~\ref{prop:reversible}.

This example suggests that some \emph{memory} is needed to implement the \spl\ and \conc\ operations so they satisfy \ref{prop:reversible} \emph{and} \ref{prop:residuals}. We show in the next section how to define this operations with a minimal amount of memory.

\subsection{Model Implementation}
\label{sec:model-implementation}

The implementation of the model is based on memory cells that store information about the events occurring at each segmentation position in a sequence.

\subsubsection{Computing the Memory Cells}

\begin{figure}[htp!]\centering
    \includegraphics[width=.6\textwidth]{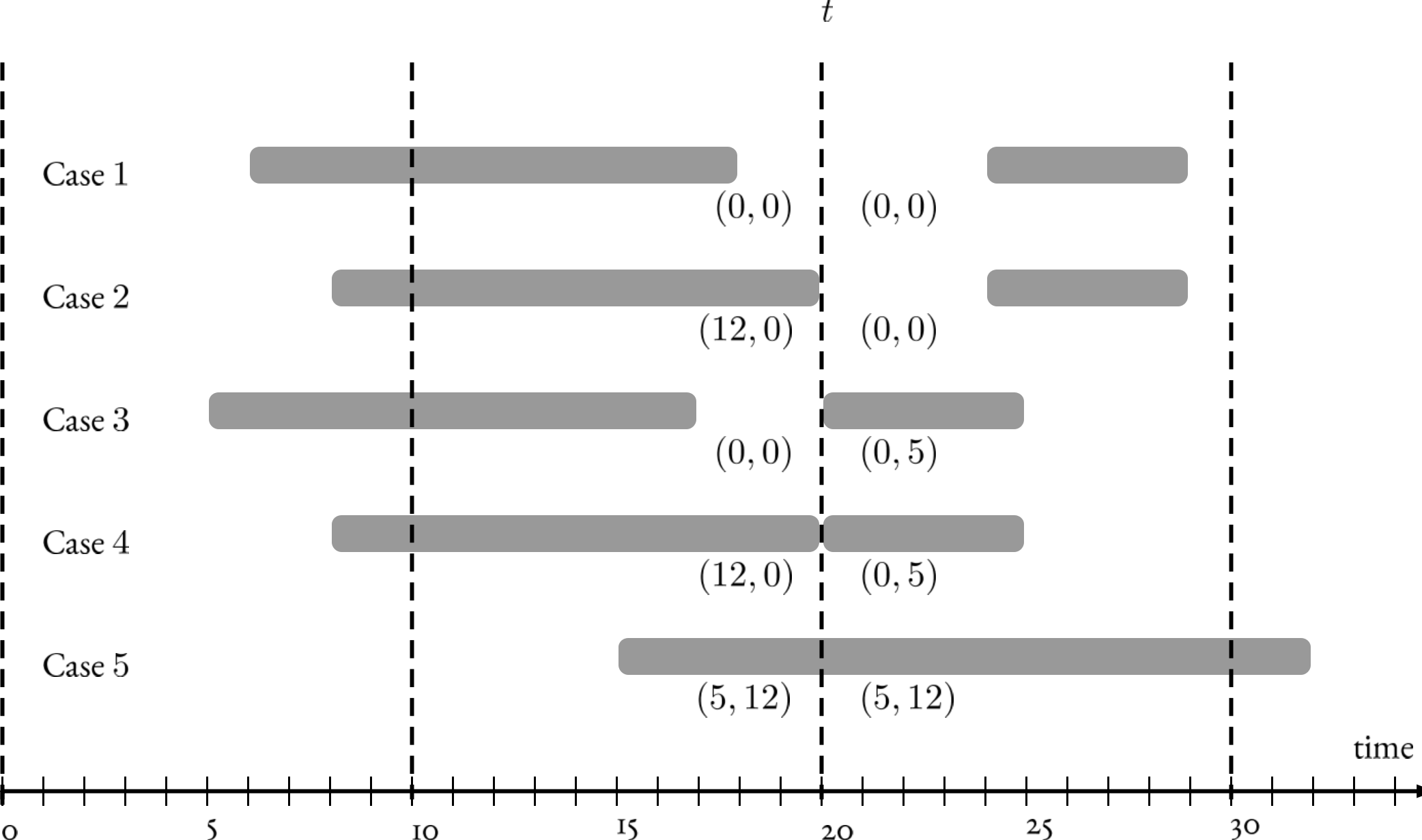}
    \caption{The left and right memory cells at segmentation position $t$. Case 1: no interval, the memory is ``empty''; case 2: a left-touching interval of length 12, stored in the left memory cell; case 3: same as 2 except on the right and length is 5; case 4: two touching intervals, each stored in the corresponding cell; case 5: a crossing interval, the left and right memories store the interval's length before and after $t$.}
    \label{fig:memory-cases}
\end{figure}

For a given segmentation position $t$, for each event $e$ containing $t$, \ie $e^- \leq t \leq e^+$, we compute the length of $e$ that lies \emph{before} $t$ and the length of $e$ \emph{after} $t$. These two quantities are stored in two memory cells, a \emph{left} and a \emph{right} memory cell, as shown in Figure~\ref{fig:memory-cases}. The left cell stores only information related to events starting strictly before $t$ and, conversely, the right cell stores information related to events ending strictly after $t$. There are five possible configurations, which are shown in Figure~\ref{fig:memory-cases}.
When the sequence is split at position $t$, the memory cells of $S$ at position $t$ are distributed to the resulting sequences: the left cell is associated to the left sequence, at position $t$ and the right cell is associated to the right sequence at position $0$. These values will be used to concatenate these subsequences with other sequences, using the \conc\ operation.
Algorithm~\ref{alg:compute_memory} computes the two memory cells for a sequence and a specific segmentation position.

\begin{algorithm}[htp]
    \caption{Compute $(l_l, l_r)$, $(r_l, r_r)$, the left and right memory cells of $S$ at $t$.}
    \begin{algorithmic}[1]
        \Procedure {ComputeMemory}{$S$, $t$}
        \If {$t = 0$}\Comment{$t$ is the start time of $S$}
            \If {$\exists e \in S$ such that $e^- = 0$} $\textbf{return } (0, 0), (0, d(e))$
            \Else {} $\textbf{return } (0, 0), (0, 0)$
            \EndIf
        \ElsIf {$t = d(S)$} \Comment{$t$ is the end time of $S$}
            \If{$\exists e \in S$ such that $e^+ = d(S)$} $\textbf{return } (d(e), 0), (0, 0)$
            \Else {} $\textbf{return } (0, 0), (0, 0)$
            \EndIf
        \Else
            \Comment{$t$ is \emph{within} the sequence}
            \If{$\exists e \in S$ with $e^- < t < e^+$} $\textbf{return }  (t - e^-, e^+ - t), (t - e^-, e^+ - t)$
            \ElsIf{$\exists e_1, e_2 \in S$ with $e_1^+ = e_2^- = t$} $\textbf{return } (d(e_1), 0), (0, d(e_2))$
            \ElsIf{$\exists e \in S$ with $e^- = t$} $\textbf{return } (0, 0), (0, d(e))$
            \ElsIf{$\exists e \in S$ with $e^+ = t$} $\textbf{return } (d(e), 0), (0, 0)$
            \Else {} $\textbf{return } (0, 0), (0, 0)$
            \EndIf
        \EndIf
        \EndProcedure

    \end{algorithmic}
    \label{alg:compute_memory}
\end{algorithm}

\subsubsection{Implementation of \spl}
\label{sec:split}

The \spl\ primitive, described by Algorithm~\ref{alg:split}, consists in distributing the intervals to the left and right sequences, $S_l$ and $S_r$, as follows: intervals occurring before the segmentation position $t$ are allocated to $S_l$ and intervals occurring  after $t$ are allocated to $S_r$. The memory of the original sequence is copied to the resulting sequences in a similar way: memory cells stored before $t$ are copied to the left sequence at the same position; memory cells of $S$ stored after $t$ are copied to the right sequence, with offset $-t$ as all sequences start at position 0 (see Section~\ref{sec:model}).

\begin{algorithm}[htp]
    \caption{Split $S$ at $t$.}
    \begin{algorithmic}[1]
    \Procedure {ComputeMemory}{$S$, $t$}

        \Comment{$\delta$ is the regular subdivision of time; $d(S)=n\delta$; $t=m\delta$ with $0 < m < n$.}

        \State $S_l \gets$ empty sequence with $d(S_l)=t$\;
        \State $S_r \gets$ empty sequence with $d(S_r)=d(S) - t$\;
        \Comment{copies the memory cells from $S$}
        \For{$i = 0, \dots, m$}
            \State $M(S_l, i\delta) \gets M(S, i\delta)$
        \EndFor
        \For{$i = m, \dots, n$}
            \Comment{translate by $-t$, as $S_r$ starts at $0$}
            \State $M(S_r, i\delta - t) \gets M(S, i\delta)$
        \EndFor
        \For{$e \in S$}
            \If{$e^+ \leq t$}
                \State \textbf{add} $e$ \textbf{to} $S_l$
            \ElsIf{$e^- \geq t$}
                \State \textbf{add} event $[e^- - t, e^+ - t]$ \textbf{to} $S_r$ \label{alg:split:add_right}
            \Else \Comment{$e$ satisfies $e^- < t < e^+$}
                \State $(l_l, l_r), (r_l, r_r) \gets M(S, t)$ \Comment{shortcut notations}
                \If{$l_l \geq \varepsilon$ \textbf{or} ($l_r = 0$ \textbf{and} $l_l > 0$)}
                    \State \textbf{add} event $[\max(0, t-l_l), t]$ \textbf{to} $S_l$ \label{alg:split:add_left_split}
                \EndIf
                \If{$r_r \geq \varepsilon$ \textbf{or} ($r_l = 0$ \textbf{and} $r_r > 0$)}
                    \State \textbf{add} event $[(0, \min(d(S), r_r)]$ \textbf{to} $S_r$ \label{alg:split:add_right_split}
                \EndIf
            \EndIf
        \EndFor
        \State \textbf{return } $S_l, S_r$ \Comment{$d(S_l)=t=m\delta$ and $d(S_r)=d(S)-t$}
    \EndProcedure
    \end{algorithmic}
\label{alg:split}
\end{algorithm}

A specific treatment is required at position $t$, to avoid creating residuals that could appear when splitting short intervals containing position $t$, in order to satisfy Property~\ref{prop:residuals}(\ref{prop:residuals:a}). If an event $e$ contains $t$, \ie $e^- < t < e^+$, we consider the memory cells $l_l$, $l_r$, $r_l$, and $r_r$ stored in $M(S, t)$ to decide whether an event of length $l_l$ should be added at the very end of the left sequence. The event is added if it is not a residual, \ie $l_l > \varepsilon$ or if it is a residual that already existed in $S$, \ie $l_r = 0$ and $l_l > 0$. A similar treatment is applied to decide if an event of length $r_r$ is inserted at the very beginning of $S_r$. Note that, doing this, we ignore the actual events, and only consider the memory. This is reflected in lines~\ref{alg:split:add_left_split} and \ref{alg:split:add_right_split} in Algorithm~\ref{alg:split}.

Sequence $S_l$ is such that the space between $t-l_l$ and $t$ is either empty of event or it contains a single event $[t-l_l,t]$, possibly left-trimmed so it does not start before $0$:
\begin{equation}\label{eq:split:inv:S1}
    e^+ \leq t-M(S_l, t).l_l, \forall e \in S_l \text{ or }[\max \{0, t-M(S_l, t).l_l\}, t] \in S_l
\end{equation}
and $S_r$ is such that the space between $0$ and $r_r$ is either empty of event or it contains a single event $[0, r_r]$, possibly right-trimmed so it does not exceed the sequence's end time:
\begin{equation}\label{eq:split:inv:S2}
    e^- \geq M(S_r, 0).r_r, \forall e \in S_r \text{ or }[0, \min \{M(S_r, 0).r_r, d(S_r)\}] \in S_r
\end{equation}
Note that the way the memory is computed, see Algorithm~\ref{alg:compute_memory}, ensures that \eqref{eq:split:inv:S1} and \eqref{eq:split:inv:S2} are satisfied by any sequence before it is split.

\subsubsection{Implementation of \conc}\label{sec:concat}

Concatenating two sequences $S_1$ and $S_2$ creates a new sequence $S$, whose duration $d(S)=d-1+d_2$, with $d_1=d(S_1)$ and $d_2=d(S_2)$. All events of $S_1$ that end strictly before $d_1$ are added to $S$, as well as all events of $S_2$ that start strictly after $0$. The only delicate operation is to decide what to do at position $d_1$ in $S$.

In Section~\ref{sec:split}, we have seen that all sequences satisfy \eqref{eq:split:inv:S1} and \eqref{eq:split:inv:S2}. Therefore, to concatenate $S_1$ and $S_2$, we need to consider four cases:
\begin{enumerate}
    \item $e^+ \leq t-M(S_1, d_1).l_l, \forall e \in S_l$ and $e^- \geq M(S_2, 0).r_r, \forall e \in S_2$, \ie $S_1$ has no event overlapping with the temporal segment defined by $M(S_1, d_1).l_l$, its left-memory at $d_1$ and, similarly, $S_2$ has no event overlapping with the temporal segment defined by $M(S_2, 0).r_r$, its right-memory at $0$;
    \item $e^+ \leq t-M(S_1, d_1).l_l, \forall e \in S_l$ and $[0, \min \{M(S_2, 0).r_r, d(S_2)\}] \in S_2$, \ie $S_1$ has no event overlapping with the temporal segment defined by $M(S_1, d_1).l_l$, its left-memory at $d_1$ and $S_2$ has an event occupying the temporal segment defined by $M(S_2, 0).r_r$, its right-memory at $0$;
    \item $[\max \{0, t-M(S_1, d_1).l_l\}, t] \in S_l$ and $e^- \geq M(S_2, 0).r_r, \forall e \in S_2$ has an event occupying the temporal segment defined by $M(S_1, d_1).l_l$, its left-memory at $d_1$ and $S_2$ has no event overlapping with the temporal segment defined by $M(S_2, 0).r_r$, its right-memory at $0$;
    \item $[\max \{0, t-M(S_1, d_1).l_l\}, t] \in S_l$ and $[0, \min \{M(S_2, 0).r_r, d(S_2)\}] \in S_2$, \ie $S_1$ has an event occupying the temporal segment defined by $M(S_1, d_1).l_l$, its left-memory at $d_1$ and, similarly, $S_2$ has an event occupying the temporal segment defined by $M(S_2, 0).r_r$, its right-memory at $0$;
\end{enumerate}
These four cases correspond to the four \textbf{if} statements in Algorithm~\ref{alg:concat} at lines~\ref{alg:concat:p1_and_p2}, \ref{alg:concat:p1_and_not_p2}, \ref{alg:concat:not_p1_and_p2}, and \ref{alg:concat:not_p1_and_not_p2} respectively.

\paragraph{Case 1.} The space delimited by the memories of $S_1$ and $S_2$ does not contain any event. The question is to decide whether an event should be created, based on the values of the memories of $S_1$ and $S_2$.

\begin{figure}[htp!]\centering
    \includegraphics[width=.7\columnwidth]{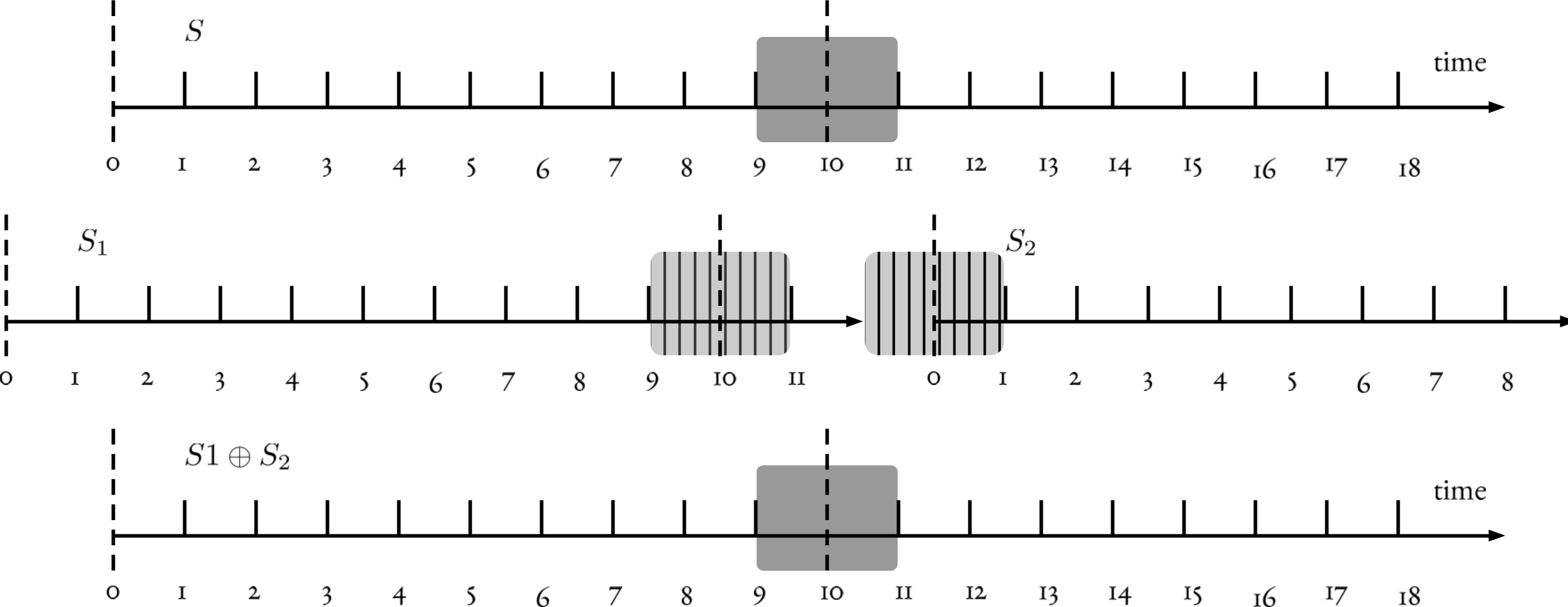}
    \caption{Splitting a sequence $S$ as $t=10$ with a residual at $t$. The resulting sequences $S_1$ and $S_2$ are empty, but memorize the residual (lightgray hatched boxes). When concatenating them, Algorithm~\ref{alg:concat} will recreate the residual.}
    \label{fig:split_concat_dancing}
\end{figure}

If the memories of $S_1$ and $S_2$ are identical, \ie $l_l=r_l$ and $l_r=r_r$, then we add the event
                \[e=[\max\{0, d_1-l_l\},\min\{d_1+d_2,d_1+r_r\}].\]
We add this event, regardless of its duration, even if it is very small, \ie $l_l+r_r<\varepsilon$, because such an short event was present at some point, before $S_1$ and $S_2$ were obtained by splitting a longer sequence. We must add this event to ensure that Property~\ref{prop:reversible} is not violated. See Figure~\ref{fig:split_concat_dancing}.

On the contrary, if the memories of $S_1$ and $S_2$ differ, the choice will depend on the values stored in the memories of $S_1$ and $S_2$. There are several cases to consider:

If $l_l=0$ or $r_r=0$, we do nothing, as no events are recorded around the concatenation position. Otherwise, that is when $l_l > 0$ and $r_r > 0$, we know that $S_1$ (resp. $S_2$) had originally an event containing position $d(S_1)$ (resp. $0$), which disappeared after a split operation. We will create an additional event
\[e=[\max\{0, d_1-l_l\}, \min\{d1+s_2, d_1+r_r\}],\]
if $d(e)=\min\{d_1+d_2, d_1+r_r\}-\max\{0, d_1-l_l\}\geq\varepsilon$, to avoid creating a residual.

\paragraph{Case 2.} The space delimited by the memories of $S_1$ and $S_2$ contains an event $e$ starting at the onset of $S_2$. The question is whether $e$ should start earlier, \ie ``in $S_1$'', depending on the memories of $S_1$ and $S_2$.

In this case, we start $e$ at \[e^-=\max\{0, d_1-l_l\},\]
another option is to set \[e^-=\max\{0, d_1-l_l, d_1 - r_l\}.\]
In the second option, the algorithm will try to create an attack for the corresponding note based on the memory of $S_2$. Both options are equally valid, the intuition is that favoring the memory of the left sequence tends to preserve temporal deviations in attacks as they appear in the left sequence. Conversely, favoring the memory of the right sequence tends to replicate deviations of attacks as they appear in the right sequence.

\paragraph{Case 3.} The space delimited by the memories of $S_1$ and $S_2$ contains an event $e$ ending at the end of $S_1$. The question is whether $e$ should end later, ``in $S_2$'', depending on the memories of $S_1$ and $S_2$.

One possible implementation is to systematically end $e$ at \[e^+=\min\{d_1+d_2, d_1+r_r\}.\]
Another option is to start the event at \[e^+=\min\{d_1+d_2, d_1+r_r, d_1+l_r\}.\]
Here again, both options are equally valid. The intuition is that favoring the memory of the left (resp. right) sequence tends to preserve temporal deviations in note durations as they appear in the left (resp. right) sequence.

\paragraph{Case 4.} The space delimited by the memories of $S_1$ and $S_2$ contains an event ending at the end of $S_1$ and another event starting at the onset of $S_2$. In this case, if $l_r>0$ and $r_l>0$, we merge the two events in a single one, otherwise, we do nothing.

\begin{algorithm}[htp]
    \caption{Concatenates $S_1$ and $S_2$.}
    \begin{algorithmic}[1]
    \Procedure {Concat}{$S$, $t$}
        \Comment{$m$ is the integer such that $d(S) = m\delta$}
        \State $d_1 \gets d(S_1)$ and $d_2 \gets d(S_2)$
        \State $S \gets$ empty sequence with $d(S) = d_1 + d_2$
        \For{$i = 0, \dots, m$}
            \Comment{Copies memory from $S_1$ and $S_2$, except at $d_1$}
            \If{$i\delta < d_1$} $M(S, i\delta) \gets M(S_1, i\delta)$ \EndIf
            \If{$i\delta > d_1$} $M(S, i\delta) \gets M(S_2, i\delta - d_1)$ \EndIf
        \EndFor
        \State $M(S, d_1).l_l \gets M(S_1,d_1).l_l$ \Comment{$l_l$ at the end of $S_1$}
        \State $M(S, d_1).l_r \gets M(S_1,d_1).l_r$ \Comment{$l_r$ at the end of $S_1$}
        \State $M(S, d_1).r_l \gets M(S_2,0).r_l$ \Comment{$r_l$ at the start of $S_2$}
        \State $M(S, d_1).r_r \gets M(S_2,0).r_r)$ \Comment{$r_r$ at the start of $S_2$}

        \For{$e \in S_1$ s.t. $e^+ < d_1$}\Comment{events ending before $d_1$ are added}
            \State \textbf{add} $e$ \textbf{to} $S$
        \EndFor
        \For{$e \in S_2$ s.t. $e^- > 0$}\Comment{events starting after $d_1$ are added}
            \State \textbf{add} event $(e^- + d_1, e^+ + d_1)$ \textbf{to} $S$
        \EndFor

        \State $(l_l, l_r), (r_l, r_r) \gets M(S, d_1)$ \Comment{shortcut notations}
        \State $P_1 \gets S_1$ is empty \textbf{or} $S_1[-1]^+ \leq d_1 - l_l$ \Comment{$S_1[-1]$ denotes the last event in $S_1$}
        \State $P_2 \gets S_2$ is empty \textbf{or} $S_2[0]^+ \geq r_r$ \Comment{$S_2[0]$ denotes the first event in $S_1$}
        \If{$P_1$ \textbf{and} $P_2$} \label{alg:concat:p1_and_p2}
            \If{$l_l = r_l$ \textbf{and} $l_r = r_r$ \textbf{and} $l_l > 0$}
                \Comment{same as initial situation for $S_1$ and $S_2$}
                \State \textbf{add} $[\max\{0, d_1 - l_l\}, \min\{d_1 + d_2, d_1 + r_r\}]$ \textbf{to} $S$ \Comment{restore memory, regardless of $\varepsilon$}
            \Else
                \If{$l_l = 0$ \textbf{and} $r_r = 0$}
                    \State do nothing \Comment{no memory to consider}
                \ElsIf{$l_l = 0$ \textbf{and} $r_r > 0$ \textbf{and} $r_l = 0$}
                    \State \textbf{add} $e=[d_1, \min\{d_1 + d_2, d_1 + r_r\}]$ \textbf{to} $S$
                    \Comment{$e\in S_2$ was not created by splitting}
                \ElsIf{$r_r = 0$ \textbf{and} $l_l > 0$ \textbf{and} $l_r = 0$}
                    \State \textbf{add} $e=[max\{0, d_1 - l_l\}, d_1]$ \textbf{to} $S$
                    \Comment{$e\in S_1$ was not create by splitting}
                \Else \Comment{ambiguous case}
                    \If{$\min\{d_1 + d_2, d_1 + r_r\} - \max\{0, d_1 - l_l\} \geq \varepsilon$}
                        \State \textbf{add} $e=[max\{0, d_1 - l_l\}, \min\{d_1 + d_2, d_1 + r_r\}]$ \textbf{to} $S$
                        \Comment{add $e$ if not residual}
                    \EndIf
                \EndIf
            \EndIf
        \ElsIf{$P_1$ \textbf{and not} $P_2$} \label{alg:concat:p1_and_not_p2}
            \State \textbf{add} $[max\{0, d_1 - l_l\}, d_1 + S_2[0]^+]$ \textbf{to} $S$ \Comment{extend $S_2[0]$ in the past (starts before $d_1$)}
        \ElsIf{not $P_1$ \textbf{and} $P_2$} \label{alg:concat:not_p1_and_p2}
            \State \textbf{add} $[S_1[0]^-, \min\{d_1 + d_2, d_1 + r_r\}]$ \textbf{to} $S$ \Comment{extend $S_1[-1]$ in the future (ends after $d_1$)}
        \Else \label{alg:concat:not_p1_and_not_p2}
            \If{$l_r > 0$ \textbf{and} $r_l > 0$}
                \State \textbf{add} $[S_1[-1]^-, d_1+S_2[0]^+]$ \textbf{to} $S$ \Comment{merge $S_1[-1]$ and $S_2[0]$}
            \Else
                \State \textbf{add} $S_1[-1]$ \textbf{to} $S$ \Comment{add both $S_1[-1]$ and $S_2[0]$}
                \State \textbf{add} $[d_1 + S_2[0]^-, d_1 + S_2[0]^+]$ \textbf{to} $S$
            \EndIf
        \EndIf
        \Return{$S$}
    \EndProcedure
    \end{algorithmic}
\label{alg:concat}
\end{algorithm}

\subsection{The Model Satisfies Properties~\ref{prop:reversible} and \ref{prop:residuals}}
\label{sec:model-discussion}

Properties~\ref{prop:reversible} and \ref{prop:residuals} respectively ensure that the model enables temporal sequence editing without creating undesired residuals and with the guaranty that editing is non-destructive. Therefore the model naturally provides an undo mechanism: if two sequences that originally formed a single sequence are concatenated together in their original position, this will result in the original sequence. But this undo mechanism is more general as even if the two sequences were used for other intermediate edit operations, it will always be possible to recreate the initial sequence. This, combined with the possibility to copy sequences, gives rise to powerful sequence edition tools. This is illustrated on \midi\ streams in Section\ref{sec:model-on-piano-rolls}.

The model relies on a memory structure computed at each segmentation position. This memory consists of two memory cells, representing the left-hand and right-hand sides of the sequence at the specified position. The essential invariant in the implementation of the \spl\ and \conc\ primitives is that
\begin{itemize}
    \item the left-hand side memory of a sequence at its end position is never modified and
    \item the right-hand side memory of a sequence at position $0$ is never modified.
\end{itemize}
Therefore, arbitrary editing the sequence never leads to any information loss, as the model always remembers information about the initial state of a sequence at its starting and end positions. This is what makes it possible to satisfy Properties~\ref{prop:reversible} and \ref{prop:residuals}.

We will not fully demonstrate that the model satisfies Properties~\ref{prop:reversible} and \ref{prop:residuals}. From the description of Algorithms~\ref{alg:split} and \ref{alg:concat}, it is clear that Property~\ref{prop:residuals} on residuals is satisfied, with the subtle case discussed in the fifth case of Algorithm~\ref{alg:concat} (see Section~\ref{sec:concat}). The Property~\ref{prop:reversible} on reversibility is easily checked in most cases. The only tricky case is when residuals are present in the edited sequences. A full proof requires reviewing all possible configurations, which would be too long. However, the discussion in Section~\ref{sec:concat} covers the most difficult case of split residuals which need to be recreated when concatenating two sequences that were originally forming a single sequence.

\subsection{Extending the Model}

The model may be extended easily to handle additional musical information. A first extension is to associate each event in a sequence to some metadata. In the case of \midi\ files, it is natural to store the velocity and the channel of a note-on event with the corresponding event. This is straightforward to implement by storing the metadata in the memory cells computed by Algorithms~\ref{alg:compute_memory}, and by associating the best metadata to newly created events in Algorithms~\ref{alg:split} and~\ref{alg:concat}.

Technically, each event $e$ is associated to a value $v(e)$, which may represent any metadata. A time event $e$ is therefore a 3-tuple $[e^-, e^+, v(e)]$. The only modifications to the algorithms are:
\begin{enumerate}
    \item in Algorithm~\ref{alg:compute_memory}, each memory cell is a 3-tuple, \eg the left memory cell for event $e$ is $[l_l, l_r, v]$;
    \item in Algorithm~\ref{alg:split}, in line~\ref{alg:split:add_right}, we add event $[e^- - t, e^+ - t, v(e)]$ instead of simply $[e^- - t, e^+ - t]$, similarly in line~\ref{alg:split:add_left_split}, we add $[e^-, t, v(e)]$ and in line~\ref{alg:split:add_right_split}, we add $[0, e^+ - t, v(e)]$;
    \item in Algorithm~\ref{alg:concat}, events receive the value stored in the corresponding memory; when we merge two events, we systematically chose to keep the metadata associated to the left event. This decision is arbitrary and we could decide otherwise, \eg use the memory of the right sequence, compute an ``average'' value.
\end{enumerate}

In our model, residuals are defined by an absolute threshold duration $\varepsilon$. It is also natural to use a relative definition for residuals. For instance, when splitting a sequence at position $t$, if an event $e$ is such that $e^-=t-a$ and $e^+=t+b$ with $b \gg a > \varepsilon$, depending on the musical context, one may want to dismiss the part of $e$ occurring before $t$ (or length $a$), although it is not a residual in the absolute sense as $a>\varepsilon$. A typical case is a measure containing a long, whole note starting slightly before the bar line. When splitting at the bar line, the head of the note should be removed, as it obviously belongs to the measure starting at $t$, not to the measure ending at $t$. It is quite easy to extend the model to handle relative residuals, although it makes Algorithms~\ref{alg:split} and~\ref{alg:concat} a bit longer, which is why it is not reported here.

Technically, the head (or tail) of a split event is considered a residual if its duration is shorter than $\varepsilon$ \emph{or} if it is shorter than a certain ratio of the length the whole event. Typical values for the ratio range from 1/10 (a fragment shorter than a third of the original event is considered too small to exist on its own) to 1/3. This modification leads to a slight increase in the complexity of  Algorithms~\ref{alg:split} and~\ref{alg:concat} to ensure that the model still satisfies Properties~\ref{prop:reversible} and \ref{prop:residuals}.

\subsection{Using the Model on Actual Music Performance}
\label{sec:model-on-piano-rolls}

As we said in Section~\ref{sec:model}, the model handles sequences of non-overlapping events, and as such is not directly applicable to \midi\ files. However, in a \midi\ file, for a given \midi\ pitch and a given \midi\ channel, the successive note-on and note-off events form a sequence of non-overlapping time intervals. Therefore, the model is applicable to \midi\ files if we treat each note (fixed pitch and channel) individually.
The targeted use of the model is to edit a \midi\ file capturing a musical performance, and therefore with non-quantized note events, but recorded with a specified tempo. The tempo may change during the piece, but we will consider it fixed here for the sake of clarity. The model is well-adapted to edit such \midi\ files with segmentation points set as a fraction of the meter, \eg one beat, one bar, half-a-beat.
The \spl\ and \conc\ primitives were implemented with this application in mind. The way we use the memory when concatenating sequences is to ensure that temporal deviations from the metrical structure of the music are preserved as much as possible, without creating disturbing residuals, and with the flexibility of a powerful undo mechanism.

We describe here  applications of our model to two real world examples.

\subsubsection{Transforming a 4/4 into a 3/4 Music Piece}

\begin{figure}\centering
    \includegraphics[width=1\textwidth, height=3.5cm]{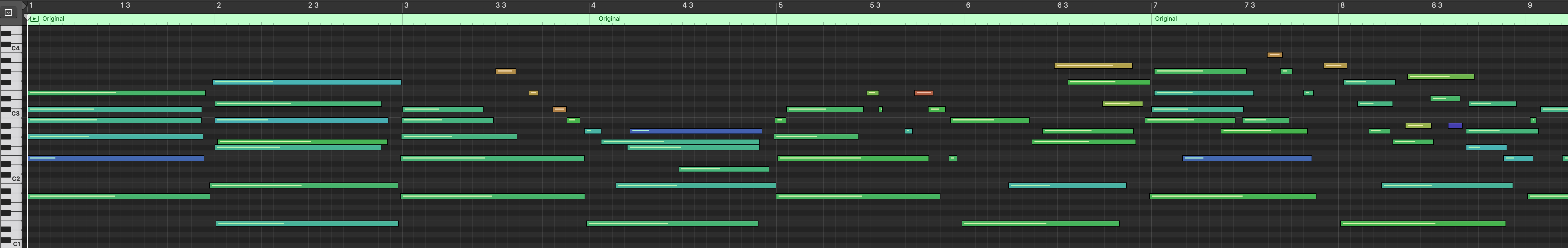}
    \caption{A piano roll showing first eight measures of a \midi\ capture of a piano performance, in the style of American jazz pianist Chick Corea.}
    \label{fig:piano-performance-original}
\end{figure}

\begin{figure}\centering
    \includegraphics[width=.75\textwidth, height=3.5cm]{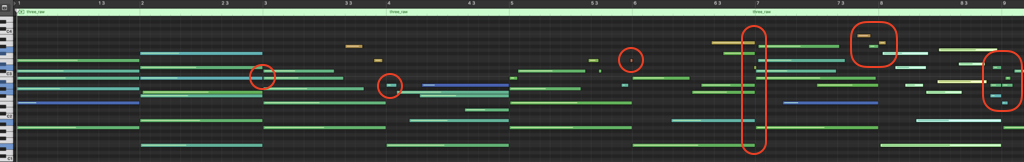}
    \caption{A \emph{raw edit} of the \midi\ shown in Figure~\ref{fig:piano-performance-original} obtained by removing the fourth beat of every measure, resulting in a version with a 3/4 time signature with many musical issues (in red boxes): some long notes are split, residuals are present, note velocities are inconsistent, and many notes are quantized.}
    \label{fig:piano-performance-3-4-raw}
\end{figure}

\begin{figure}\centering
    \includegraphics[width=.75\textwidth, height=3.5cm]{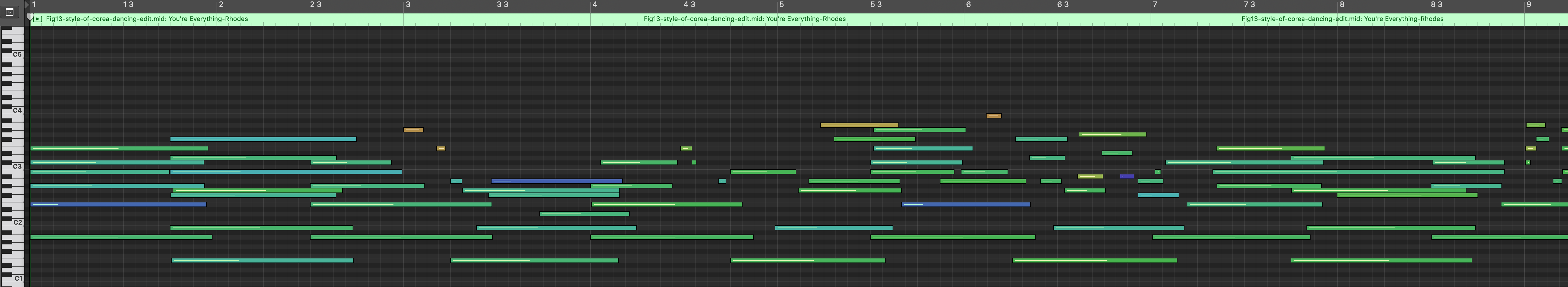}
    \caption{A Dancing \midi\ edit of the \midi\ shown in Figure~\ref{fig:piano-performance-original} obtained by removing the fourth beat of every measure, resulting in a version with a 3/4 time signature with none of the musical issues mentioned (compare white boxes here with red boxes in Figure~\ref{fig:piano-performance-3-4-raw}). Many notes are held across segmentation points, creating a smoother musical output.}
    \label{fig:piano-performance-3-4-dancing}
\end{figure}

Consider the piano roll in Figure~\ref{fig:piano-performance-original}, which shows eight measures of a \midi\ capture of a piano performance by French pianist Lionel Gaget, in the style of American pianist Chick Corea. The time signature of this piece is 4/4. We aim at producing a 3/4 version of this piece by removing the last beat of every measure.
We created two versions: Figure~\ref{fig:piano-performance-3-4-raw} shows a piano roll of the resulting \midi\ stream when performing raw edits and Figure~\ref{fig:piano-performance-3-4-dancing} shows the resulting music when performing edits with our model.

The version in Figure~\ref{fig:piano-performance-3-4-raw} exhibits many musical issues that make it necessary to manually edit the resulting piano roll to preserve the groove of the original music. Here is a list of some of these issues:
\begin{itemize}
    \item Split long notes, such as the B2 between measures 2 and 3;
    \item Residuals: A2 beginning of m.~4, B2 beginning of m.~5, loud E3 at end of m.~5, etc.
    \item Quantized notes: heads of B1 and F$\sharp$3 beginning of m.~2, tails of E1 and B1 at end of m.~6, etc.
    \item Inconsistent velocities: the velocity of the split B2 changes between m.~2 and m.~3.
\end{itemize}

The piano roll in Figure~\ref{fig:piano-performance-3-4-dancing}, on the other hand, does not have any of the issues of the raw edit version, and, as a result, sounds more natural and preserves the style of original music more convincingly.

\subsubsection{Harmonization}

Recent progress in Artificial Intelligence have led to powerful music generation systems, especially in the symbolic domain \cite{briot2017deep}. Computers have become extremely efficient at creating brief musical fragments in many styles \cite{VanDenOord:2016:arXiv:WaveNet,Hadjeres:2017:ICML:Deepbach}, but no algorithm has yet captured the art of spontaneously arranging musical material into longer convincing structures, such as songs. For instance, in the composition process using \textsc{FlowMachines} \cite{Papadopoulos:2016:CP:FlowComposer}, human musicians are in charge of selecting and organizing musical material suggested by the computer to create large-scale structures conveying a sense of direction. These new ways of making music highlight the need for powerful tools to edit musical material.

\begin{figure}\centering
    \includegraphics[width=.9\textwidth]{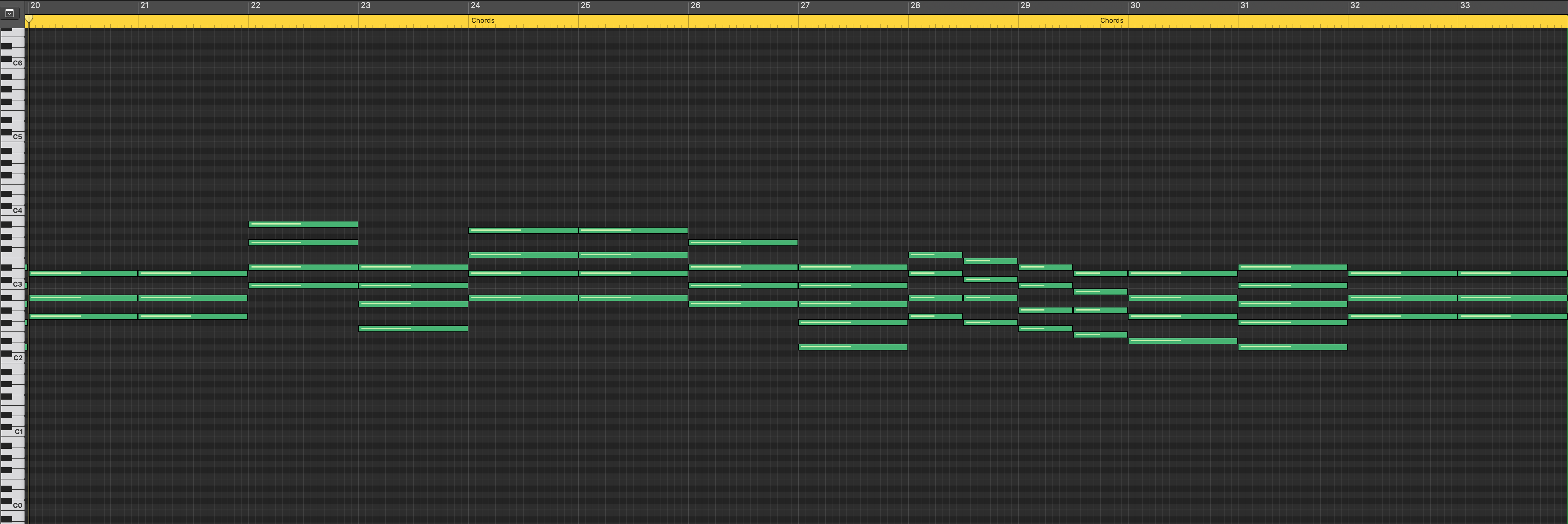}
    \caption{The harmonization of ``Autumn Leaves'' as written in the lead sheet from measure 20 to measure 33. The list of chords visible on the figure is Gm (measures~20-21), Cm7, F7, B$\flat$ maj7 (mm.~24-25), Am7/$\flat$5, D7/9, Gm7, F$\sharp$7, Fm7, E7, E$\flat$ maj7, D7/$\flat$9, and Gm (mm.~32-33) using a jazz notation for chord symbols, in a default voicing.}
    \label{fig:autumn-chords}
\end{figure}

One of the main strength of the \midi\ format is that it enables musical transformations, such as pitch-shifting (or  \emph{transposition}) and time-stretching, with no loss of quality. The model is naturally compatible with transpositions because the memory cells do not depend on the actual pitch of a note. This can be illustrated by integrating our model within any generative algorithm.
As an illustration, we have developed a system that produces harmonizations for a given \emph{target} \midi\ file in the style of a \emph{source} \midi\ file. The harmonizer outputs a new \midi\ file with the same duration as the target file, created by editing the source file, using \spl\ and \conc\ combined with chromatic transpositions, in such a way that the output file's harmony matches that of the target \midi\ file. The harmonizer uses Dynamic Programming \cite{Bellman:1957} or similar techniques, such as Belief Propagation \cite{Papadopoulos:2015:BeliefPropagation}, to produce a file that optimizes the harmonic similarity with the target file and, at the same time, minimizes the edits in the source file, to preserve the style of the source as much as possible. We do not fully describe the harmonizer in this article \cite{Roy:2019:MIDIHarmonizer}.

\begin{figure}\centering
    \includegraphics[width=.9\textwidth]{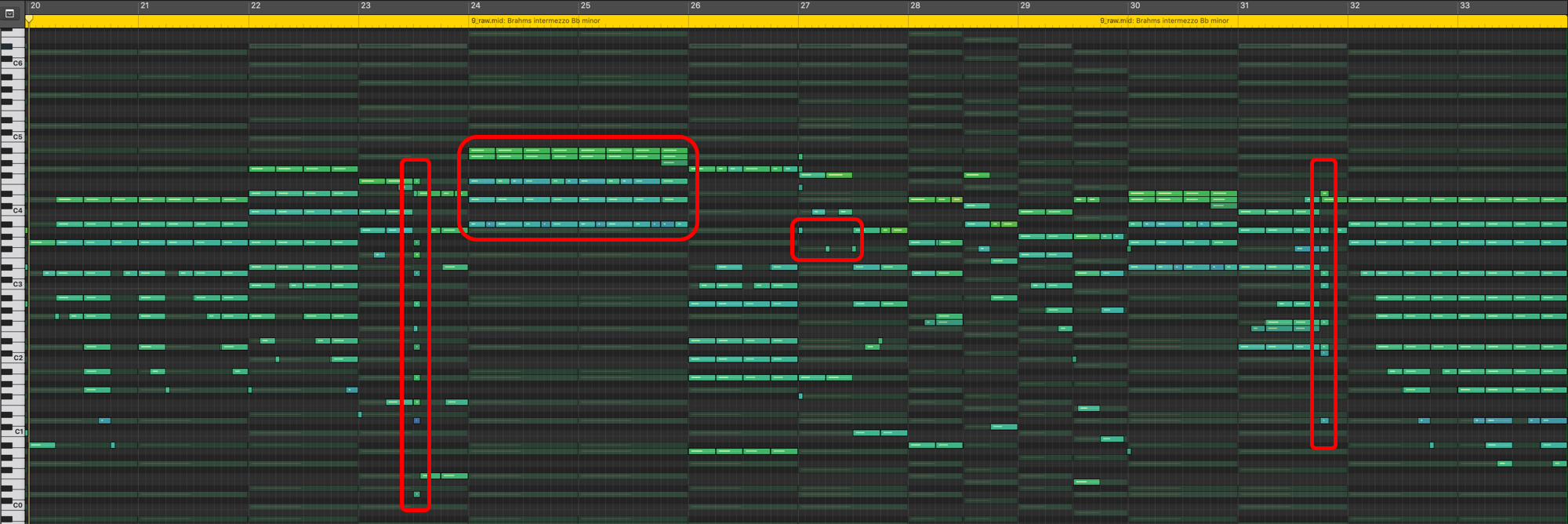}
    \caption{Harmonization of the 14 measures of ``Autumn Leaves'' shown in Figure~\ref{fig:autumn-chords} in the style of a piano performance of an intermeeezo by Johannes Brahms and using raw \midi\ edits. The actual harmony is visible in the background (light green notes). Musical inconsistencies are indicated by red boxes.}
    \label{fig:autumn-brahms-raw}
\end{figure}

\begin{figure}\centering
    \includegraphics[width=.9\textwidth]{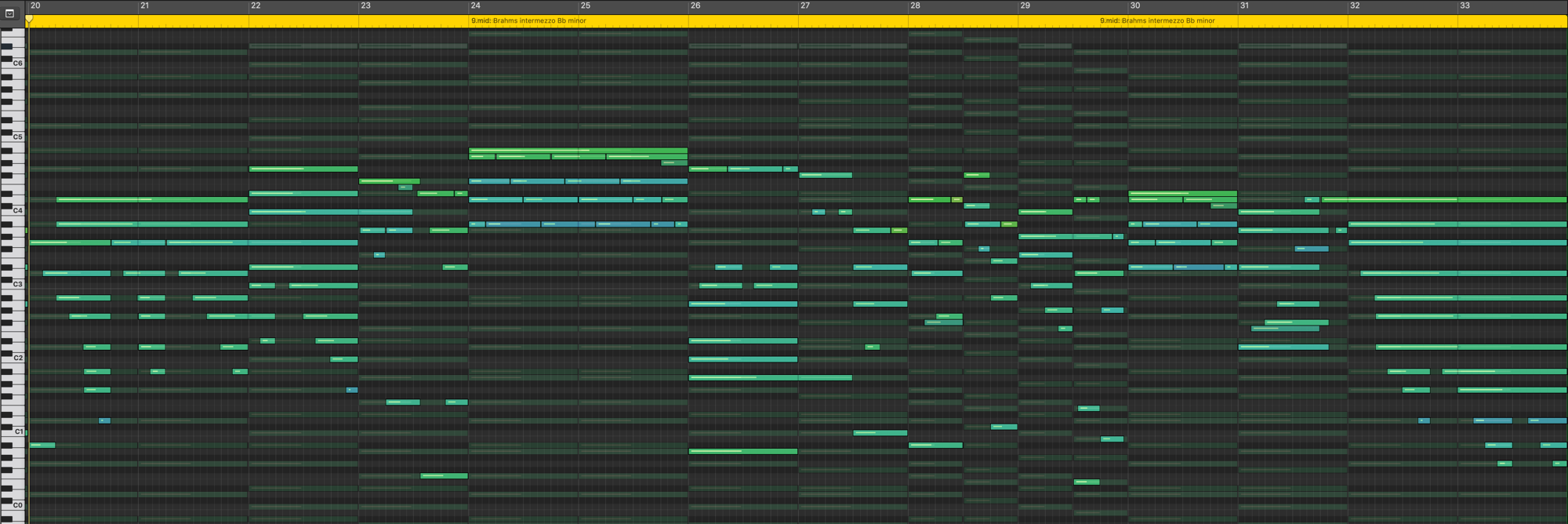}
    \caption{The same harmonization as in Figure~\ref{fig:autumn-brahms-raw} except all edits are performed with the model. This piano-roll has none of the musical issues of the piano roll in Figure~\ref{fig:autumn-brahms-raw}.}
    \label{fig:autumn-brahms-dancing-midi}
\end{figure}

Figures~\ref{fig:autumn-brahms-raw}-\ref{fig:autumn-brahms-dancing-midi} show piano rolls obtained by harmonizing 14 measures of ``Autumn Leaves'' (see Figure~\ref{fig:autumn-chords}) using a \midi\ capture of a piano performance of an intermezzo by Johannes Brahms. The raw edit in Figures~\ref{fig:autumn-brahms-raw} exhibits the same type of artifacts as Figure~\ref{fig:piano-performance-3-4-raw}. Long notes are split at each segmentation point (here, every beat), which creates dissonant chords at measures 24-25. More generally, the style of the resulting \midi\ file is substantially different from that of the original source. The source contains numerous long-held notes and most chords are arpeggiated. In the \midi\ file on Figures~\ref{fig:autumn-brahms-raw} long-held notes are broken at every beat, making most chords non-arpeggiated (notes are played all together). Many residuals are introduced, and some of them make up chords (red boxes in measure 23 and 31) that sound  wrong. On the contrary, Figure~\ref{fig:autumn-brahms-dancing-midi} shows a much cleaner result, with none of these musical issues.

\section{Conclusion}

We have presented a model for editing non-quantized, metrical musical sequences represented as \midi\ files.
We first listed a number of problems caused by the use of naive edition operations applied to performance data, using a motivating example.
We then introduced a model, called Dancing \midi\, based on 1) two desirable, well-defined properties for edit operations and 2) two well-defined operations, \spl\ and \conc, with an implementation.
We showed that our model formally satisfies the two properties, and that our model does not create most of the problems that occur with naive edit operations on our motivating example, as well as on a real-world example using an automatic harmonizer.
Our approach has limitations. First, the model requires two parameters ($\varepsilon$ and the relative ratio), which have to be set by the user. Assigning these parameters to 0.15 beats and 20\% respectively turned out to work well for most music we had to deal with. However, there are cases where these parameters should be tuned, especially for non typical music (e.g. extreme tempos or very dense).
More generally, the model is monophonic by nature, so it does not make any inference on groups of notes (e.g. chords). This may produce, in rare cases, strange behavior (like treating one note of a chord differently than others).

However the reversibility of the operations (no undo mechanism is required by construction) and the light weight nature of the model (complexity is linear in the number of segmentation points) makes it worth using in many cases as it clearly improves on naive editing.

\section*{acknowledgement}
We thank Jonathan Donier for fruitful remarks and comments.

\bibliographystyle{acm}
\bibliography{dancing-midi}
\end{document}